\documentclass{ws-rv9x6}
\usepackage{ws-rv-van}
\pdfoutput=1
\makeindex

\begin{document}

\chapter[Nonlinear pulse propagation in a disordered medium ]{An effective theory of pulse propagation in a nonlinear and disordered medium in two dimensions}
\author{Georg Schwiete}
\address{Department of Physics, Texas A\&M University, College Station, TX 77843-4242, USA}
\author[G. Schwiete and A. M. Finkel'stein]{Alexander Finkel'stein}
\address{Department of Physics, Texas A\&M University, College Station, TX 77843-4242, USA\\ and Department of
Condensed Matter Physics, The Weizmann Institute of Science, 76100
Rehovot, Israel}

\begin{abstract}
We develop an effective theory of pulse propagation in a nonlinear {\it and} disordered medium. The theory is formulated in terms of a nonlinear diffusion equation. Despite its apparent simplicity this equation describes novel phenomena which we refer to as "locked explosion" and "diffusive" collapse. The equation can be applied to such distinct physical systems as laser beams propagating in disordered photonic crystals or Bose-Einstein condensates expanding in a disordered environment.
\end{abstract}

\body

\section{Introduction}

\begin{figure}[t!]%
\begin{center}
  \parbox{2.1in}{\includegraphics[width=1.9in,angle=270,clip=]{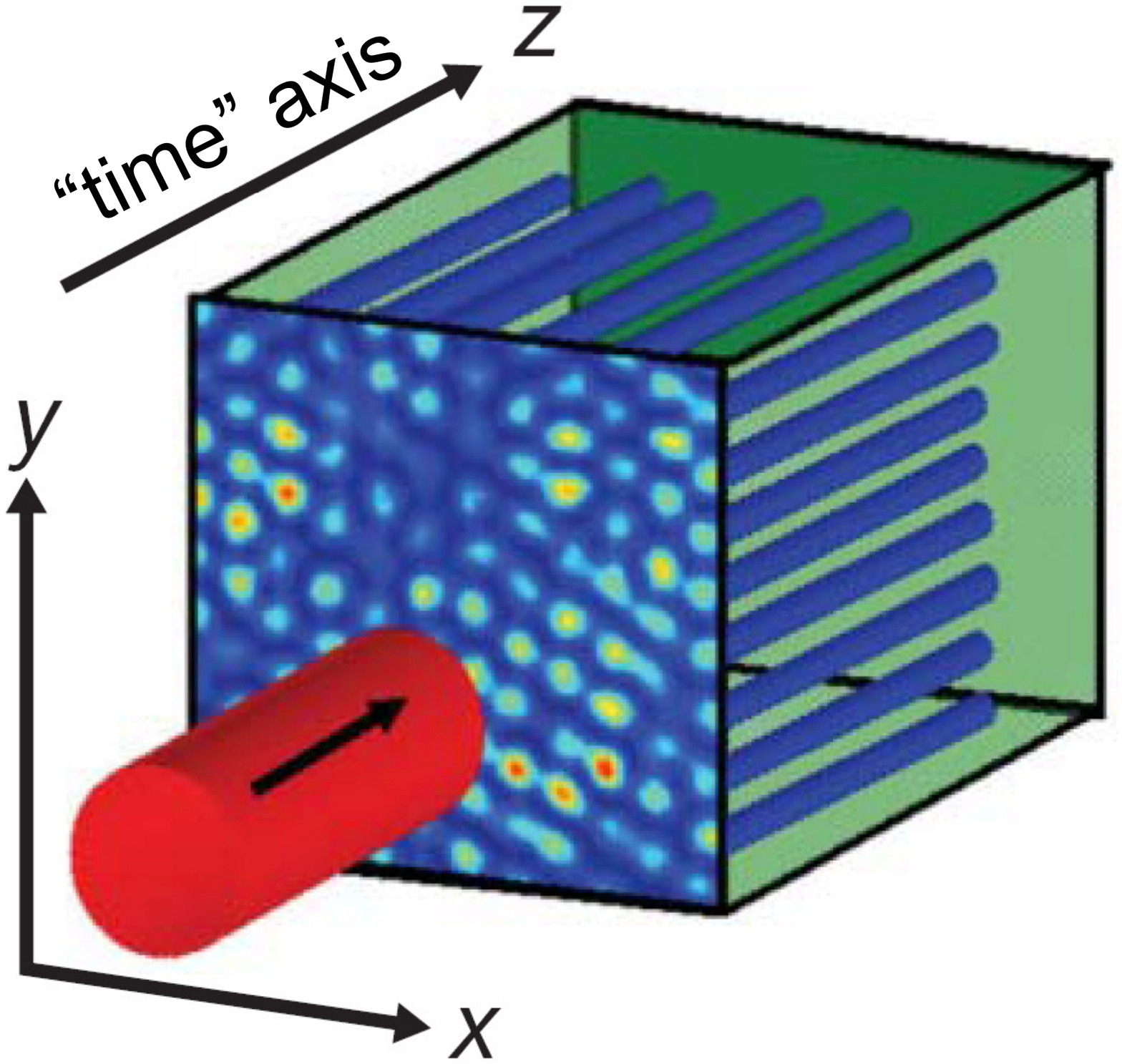}
  \figsubcap{a}}
  \hspace*{4pt}
  \parbox{2.1in}{\includegraphics[width=1.9in,angle=270,clip=]{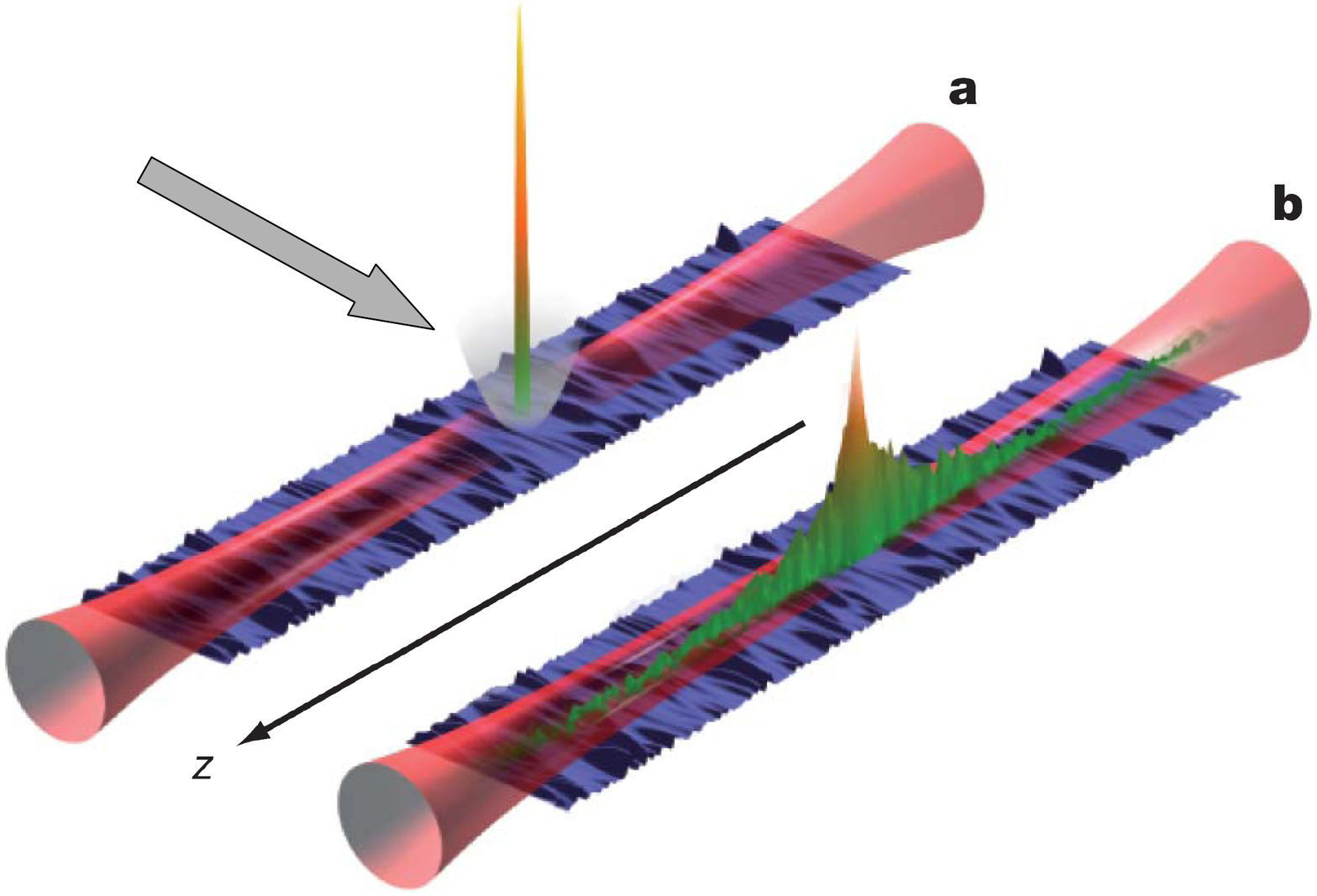}
  \figsubcap{b}}
  \caption{
  (a) In the experiment of Ref.~\refcite{Schwartz07} a laser beam (in red) is sent onto a disordered nonlinear photonic crystal. Inside the crystal, the evolution of the smooth envelope of the electric field in the transverse (xy) plane is approximately described by the nonlinear Schr\"odinger equation (NLSE). In this equation, the $z$-coordinate plays the role of "time". The intensity profile of the outgoing beam is measured at the opposite side of the crystal. Adapted from Schwartz et al., Ref.~\refcite{Schwartz07}.
  (b) In the experiment of Ref.~\refcite{Billy08} an atomic Bose-Einstein condensate is released from a small trap (displayed in white) and subsequently expands along a one-dimensional channel, formed by an external confining potential (in red). During the expansion, the atoms are subjected to a disorder potential (in blue). After a while, the motion comes to a halt and the condensate becomes localized. The experiment was performed at small densities, so that the nonlinear term in the Gross-Pitaevskii equation can be considered as small. Adapted from Billy et al., Ref.~\refcite{Billy08}.}
  \label{fig:experiments}%
\end{center}
\end{figure}

\begin{figure}[b!]
\centerline{\includegraphics[width=5cm,angle=270,clip=]{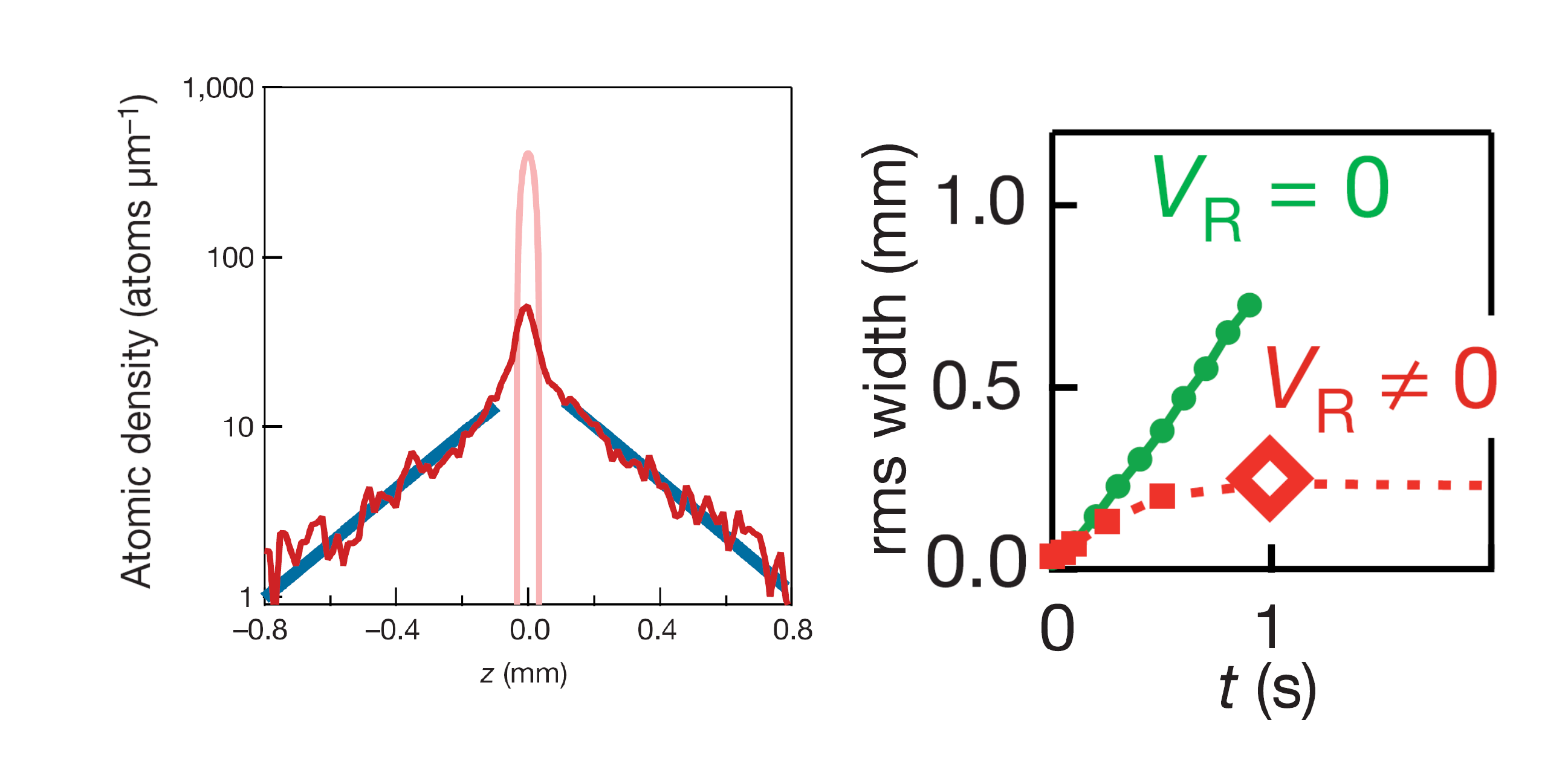}}
\caption{On the left hand side, the profile of the condensate at the final stage of the experiment of Ref.~\refcite{Billy08} is displayed on a logarithmic scale. The tails of the wave function decay exponentially. This is interpreted as evidence for Anderson localization. On the right hand side the rms width of the condensate is plotted as a function of time. In the presence of a disorder potential (red line), the width approaches a constant value, while in the absence of disorder (green line), the condensate expands ballistically. Adapted from Billy et al., Ref.~\refcite{Billy08}.}
\label{fig:aspectprofile2}
\end{figure}

\begin{figure}
\centerline{\includegraphics[width=4.0cm,angle=270,clip=]{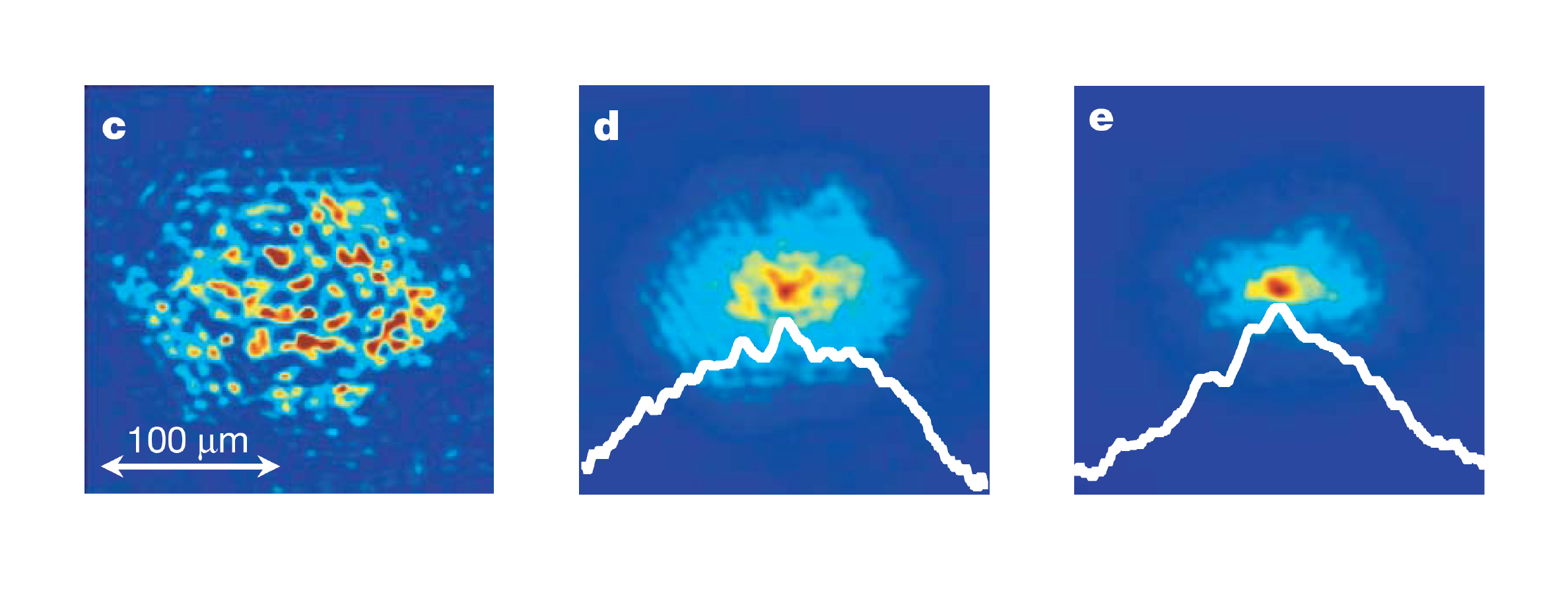}}
\caption{The intensity distribution of the outgoing laser beam in the experiment of Ref.~\refcite{Schwartz07} is shown in a regime for which the nonlinearity is negligible. The disorder level increases from left to right, starting from the clean lattice. All distributions are averaged over different disorder configurations. The white lines display the logarithm of the intensity as a function of the transverse coordinate. The authors of Ref.~\refcite{Schwartz07} interpret the distribution in the middle as the result of diffusion, and the distribution on the right as a signature of Anderson localization. Courtesy of Schwartz et al., Ref.~\refcite{Schwartz07}.}
\label{fig:segevlinear}
\end{figure}

\begin{figure}
\centerline{\includegraphics[width=5.5cm,angle=270,clip=]{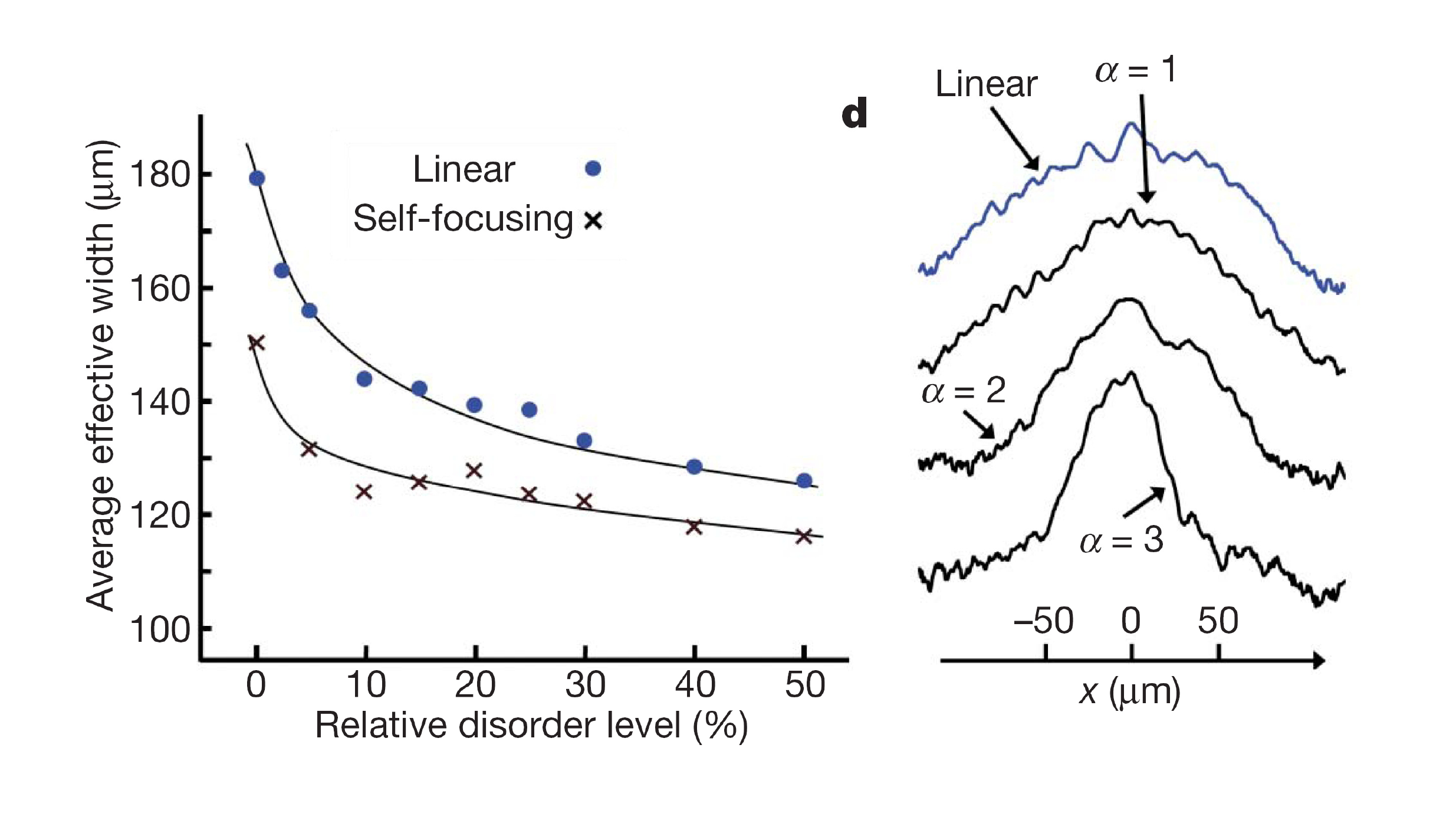}}
\caption{The average width of the outgoing laser beam as a function of the disorder level in the experiment of Ref.~\refcite{Schwartz07} is compared for the linear case and in the presence of a self-focusing nonlinearity (left panel). The right panel compares (the logarithm of) the averaged intensity profile for a fixed disorder strength as a function of the transverse coordinate. The parameter $\alpha$ is a dimensionless measure for the strength of the nonlinearity. The main result is that the self-focusing nonlinearity promotes localization. Courtesy of Schwartz et al., Ref.~\refcite{Schwartz07}.}
\label{fig:segevnonlinear1}
\end{figure}

In recent years, novel experimental techniques made possible first observations of wave-packets evolving in the presence of random scatterers and nonlinearities. In a number of optical experiments, a laser beam was sent into a nonlinear optical medium with a random refractive index, and the beam profile in the transverse direction(s) was monitored on the opposite side of the sample \cite{Schwartz07,Lahini08}, for an illustration see Fig.~\ref{fig:experiments}(a). In a second class of experiments, atoms forming a Bose-Einstein condensate were released from a trap and subjected to a disorder potential during the expansion \cite{Clement05,Fort05,Lye05,Schulte05,Billy08,Roati08}, see Fig.~\ref{fig:experiments}(b). The experiments were inspired by the idea that in these setups, unlike for transport experiments in electronic systems, one can {\it visualize} the phenomenon of Anderson localization, whereby a wave-packet or quantum particle is confined within a finite volume as a result of multiple scattering on a random potential (Figs.~\ref{fig:aspectprofile2},\ref{fig:segevlinear},\ref{fig:segevnonlinear1}).
The evolution of the injected wave-packet in both experiments can be described by the non-linear Schr\"odinger equation (NLSE), in the context of atomic Bose-Einstein condensates referred to as the Gross-Pitaevskii Equation (GPE). This equation differs from the linear Schr\"odinger equation by an additional cubic term and is used as a paradigmatic description for nonlinear waves. The nonlinearity is a consequence of interactions between particles in the case of atomic condensates and of a change in the refractive index in response to the electric field (Kerr effect) in the case of laser beams.

Motivated by these experiments we derive, starting from the GPE/NLSE, a kinetic equation that describes the evolution of an injected wave-packet in a weakly disordered nonlinear medium in two dimensions. Analysis of this equation reveals a rather nontrivial picture: Irrespective of the sign of the nonlinearity the mean square radius of the wave-packet changes linearly in time, $\partial_t\left\langle r^2\right\rangle\propto E_{tot}$, where $E_{tot}$ is the total energy of the wave-packet. For a repulsive nonlinearity the initial change of the profile displays features of an explosion, although the overall size of the wave-packet is growing slowly as in ordinary diffusion. For an attractive nonlinearity, the radius can either grow or decrease, depending on the sign of $E_{tot}$. In particular, for $E_{tot}<0$ we predict a slow "diffusive" collapse as the radius of the wave-packet shrinks towards zero.

In this paper we will mostly use the language related to the GPE, but also indicate below how to translate to a language more suitable for optical experiments. The GPE describes the evolution of a wave-function $\Psi$ \cite{Pitaevskii03}:
\begin{eqnarray}
i\partial_t\Psi({\bf r},t)=-\frac{1}{2m}\nabla^2\Psi({\bf r},t)+u({\bf r})\Psi({\bf r},t)+\lambda |\Psi({\bf r},t)|^2\Psi({\bf r},t),
\label{Eq:Gross}
\end{eqnarray}
where we set $\hbar=1$. For positive (negative) $\lambda$ this equation contains a repulsive (attractive) self-consistent potential $\lambda |\Psi({\bf r},t)|^2$. This corresponds to a nonlinearity of the de-focusing (self-focusing) type. The static disorder potential $u({\bf r})$ is the source of randomness in the equation. For simplicity we choose for our calculation a Gaussian white noise potential with correlation function $\left\langle u({\bf r})u({\bf r'})\right\rangle=\delta({\bf r}-{\bf r'})/(m\tau)$ [for a discussion of averaging for speckle potentials see, e.g., Ref.~\refcite{Kuhn07}]. The angular brackets denote averaging over disorder configurations and $\tau$ is the scattering time.

The NLSE used in optics is derived in the so-called paraxial approximation \cite{Shen84}, and describes the evolution of the smooth envelope of the electric field. The main propagation direction of the laser beam, say the $z$-direction, plays the role of time in the NLSE, see Fig.~\ref{fig:experiments}(a). In this sense, the disorder potential which results from random variations of the refractive index is static when it is $z$-independent. For example, the two-dimensional (2d) transverse evolution of a pulse is studied in a 3d sample. The intensity of the beam is proportional to $|\Psi({\bf r},z)|^2$. In the NLSE, the mass $m$ in the GPE is replaced by the wave vector $k=\omega/c$, where $\omega$ is the frequency of the carrier wave and $c$ the velocity of light in the medium.

For a condensate released from a confining harmonic oscillator potential, as is typical for experiments on cold atomic gases, the GPE \emph{without disorder} can be solved exactly \cite{Castin96,Kagan96}. During an initial stage the potential energy originating from the nonlinearity is almost entirely converted into kinetic energy. This period of violent acceleration is followed by a second stage, during which the nonlinearity is no longer essential. Expansion in the presence of disorder in the two-dimensional case was recently addressed in reference \cite{Shapiro07}. In this paper it has been assumed that for repulsive nonlinearity an initial ballistic stage is not affected by disorder, while the subsequent diffusive expansion is not affected by the nonlinearity, thereby separating the two effects. In contrast, we are interested here in the {\it interplay} of disorder and nonlinearity, both attractive and repulsive. This is especially interesting in 2d, as it is known that for linear wave propagation and weak disorder there is an extended diffusive regime preceding localization on (exponentially) large length scales. It is this regime that we address \cite{Garcia09}.

Since the system is far out of equilibrium, we choose to work with a kinetic equation. The derivation of the kinetic equation proceeds as follows. We use methods of classical statistical field theory to derive a functional integral expression for the disorder averaged density \cite{Martin73,Kamenev05}. The formalism involves a doubling of the degrees of freedom, similar to the Keldysh or closed-time-path approaches for quantum systems \cite{Kamenev05}, where two fields are introduced on forward and backward time-contours. Instead of averaging over a statistical ensemble in the initial state, we assume that the wave-function at the initial time is known. Averaging is performed over disorder configurations. Scattering on impurities is included on the level of the self-consistent Born approximation. While interference (weak localization) corrections are not covered by this approximation, it allows for a consistent description of diffusion in the presence of nonlinearity. The nonlinearity is treated by introducing a self-consistent potential $\vartheta({\bf r},t)$. In this way it is possible to include interaction effects in a non-perturbative way, which is crucial for the problem at hand. To obtain the kinetic equation for the density in the diffusive limit, we assume that the initial wave-function sets a momentum scale $p_0$ characterizing the main part of the momentum distribution, so that the weak disorder condition $p_0l\gg 1$ is fulfilled, where $l=p_0\tau/m$ is the mean free path. We further assume that the density varies smoothly on scales of $l$, in particular that the size of the condensate is much larger than the mean free path. Both of these conditions can be met simultaneously. The phase of $\Psi$, which is related to the momentum, may change rapidly, while the amplitude, which determines the density, may vary smoothly. Even if the density does not satisfy the smoothness condition initially, it is natural to expect that in the case of an expansion it will become sufficiently smooth after some time \cite{approximations}. The derivation of the kinetic equation will be presented elsewhere \cite{Schwiete09}.

\begin{figure}
\centerline{\includegraphics[width=6cm]{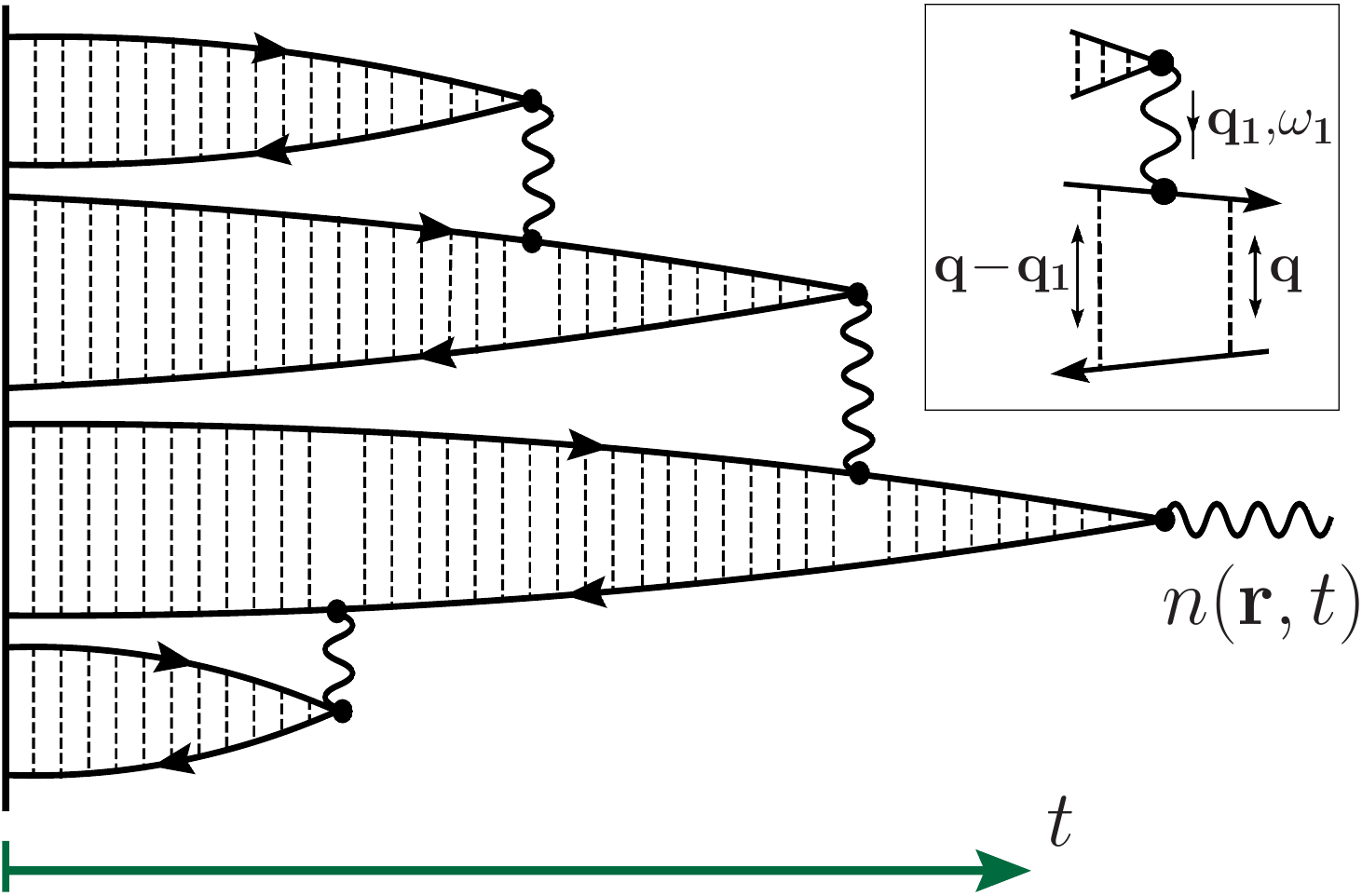}}
\caption{Graphical illustration of the kinetic equation, Eq. (\ref
{Eq:basic1}). The injection process takes place on the left. Solid lines are disorder averaged particle propagators. Ladders graphically represent density diffusion. The particular way of disorder averaging is justified for $\varepsilon_{kin}\tau\gg 1$; $\varepsilon_{kin}$ is defined below Eq.~(\ref{Eq:basic3}). The wavy lines account for the nonlinearity in Eq.~(\ref{Eq:Gross}). The block magnified in the inset gives rise to the terms $\nabla(D_\vartheta\nabla \tilde{n}({\bf r},t,\varepsilon))$ and $\partial_t\vartheta({\bf r},t)\partial_\varepsilon \tilde{n}({\bf r},t,\varepsilon)$ in the kinetic equation, Eq.~2.
}
\label{fig:diagram}
\end{figure}

\begin{figure}
\centerline{\includegraphics[width=5.5cm,angle=270]{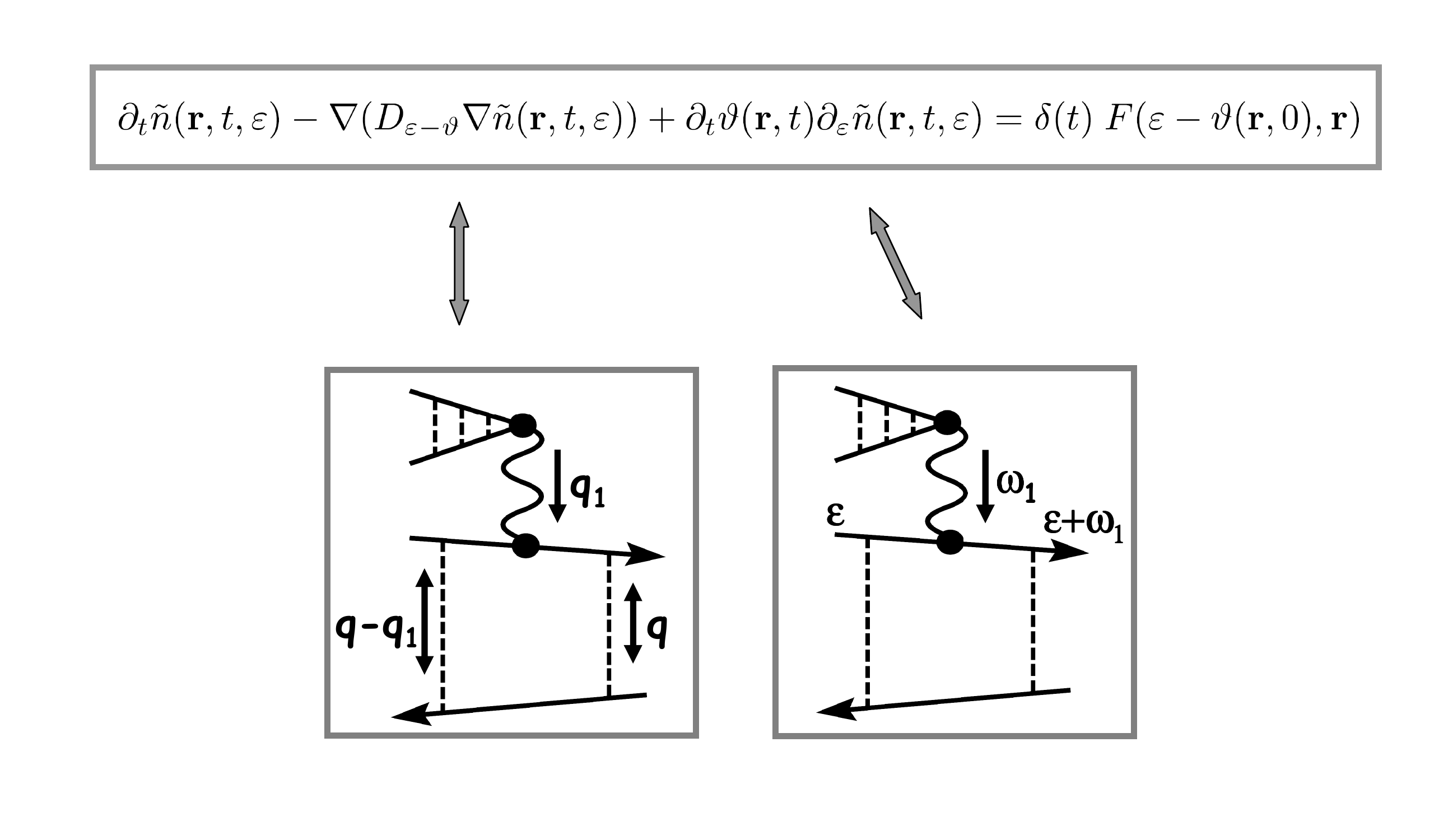}}
\caption{This figure illustrates the correspondence between different terms in the kinetic equation and a certain block in perturbation theory.}
\label{fig:blocksvseq}
\end{figure}

Starting from Eq.~(\ref{Eq:Gross}), the outlined steps lead to the following kinetic equation in the diffusive regime
\begin{eqnarray}
&&\partial_t\tilde{n}({\bf r},t,\varepsilon)-\nabla(D_{\varepsilon-\vartheta}\nabla \tilde{n}({\bf r},t,\varepsilon))+\partial_t\vartheta({\bf r},t)\partial_\varepsilon \tilde{n}({\bf r},t,\varepsilon)\nonumber\\
&&=\delta(t)\;F(\varepsilon-\vartheta({\bf r},0),{\bf r})\label{Eq:basic1}
\end{eqnarray}
where
\begin{eqnarray}
F(\varepsilon,{\bf r})=\int \frac{d^2qd^2p}{(2\pi)^4}\;F({\bf p},{\bf q})\;\exp(i{\bf q}{\bf r})\;2\pi \delta(\varepsilon-\varepsilon_{\bf p}),
\end{eqnarray}
and $F({\bf p},{\bf q})=\Psi_0({\bf p}+{\bf q}/2)\Psi_0^*({\bf p}-{\bf q}/2)$ is determined by the initial wave function $\Psi_0$; $\varepsilon_{\bf p}=p^2/(2m)$ is the kinetic energy, and $D_{\varepsilon}=\varepsilon\tau/m$ the diffusion coefficient. The equation should be supplemented with the self-consistency relation for the potential $\vartheta({\bf r},t)=2\lambda n({\bf r},t)$, where $n({\bf r},t)=\int d\varepsilon/(2\pi)\; \tilde{n}({\bf r},t,\varepsilon)$. Despite its apparent simplicity it is a rather complicated nonlinear integro-differential equation. The equation effectively sums an infinite series of diagrams of the type shown in Fig.~\ref{fig:diagram}. It is a peculiarity of the perturbation theory for a classical field equation such as the GPE that no closed loops arise\cite{Zinn02}, making it quite distinct from the related problem in interacting electron systems. The relation between certain blocks appearing in diagrammatic perturbation theory and the corresponding terms in the kinetic equation is visualized in Fig.~\ref{fig:blocksvseq}.

\begin{figure}
\centerline{\includegraphics[width=9.5cm,clip=]{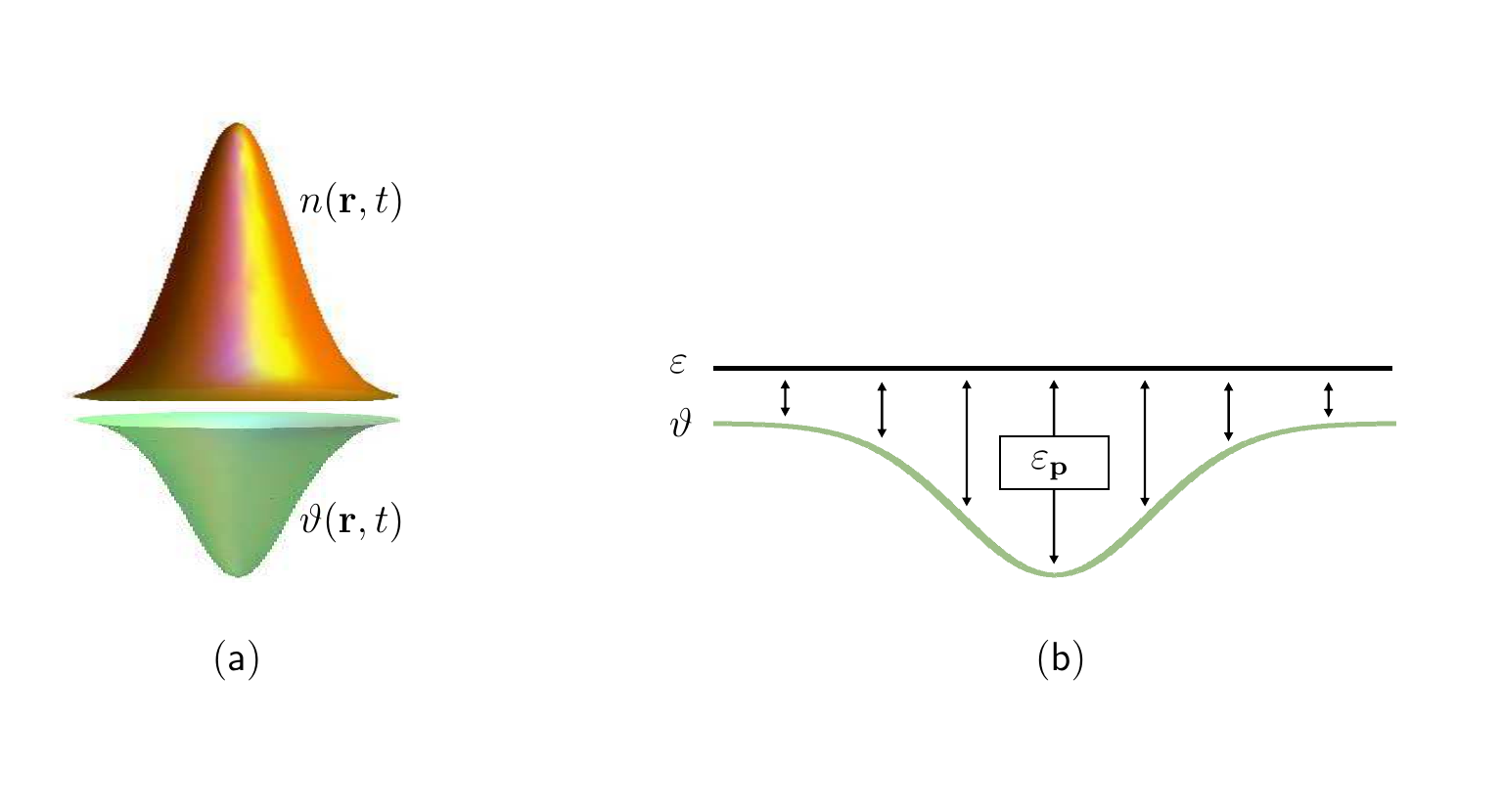}}
\caption{The density distribution $n({\bf r},t)$ creates its own self-consistent potential $\vartheta({\bf r},t)$. We illustrate the attractive (focusing) case. In a classical analogy, the variable $\varepsilon$ in Eq.~\ref{Eq:basic1} can be interpreted as the total energy and $\vartheta$ as the potential energy of a diffusing particle. Correspondingly, the diffusion coefficient is determined by the kinetic energy $\varepsilon_{\bf p}=\varepsilon-\vartheta$. }
\label{fig:classicalfigure}
\end{figure}

The physics described by this equation is essentially classical. Imagine first that the potential $\vartheta$ does not depend on time. Consider now a particle diffusing with total energy $\varepsilon$ on the  background of a smoothly varying potential $\vartheta$, see Fig.~\ref{fig:classicalfigure} for illustration. If scattering events are frequent enough, the diffusion coefficient is determined by the kinetic energy $\varepsilon_{\bf p}=\varepsilon-\vartheta$ that varies locally in space. If the potential additionally varies in time, the particle may change its total energy. If on the other hand the potential depends on time \emph{only}, the kinetic energy does not change. Therefore, it is expected that a purely time-dependent potential has no effect on the density. This observation is related to the fact that in the original GPE a purely time dependent potential $V(t)$ may be removed by a suitably chosen gauge-transformation,
\begin{eqnarray}
\Psi({\bf r},t)\rightarrow \Psi({\bf r},t)\;\exp\left(-i\int_{t_0}^t dt' V(t')\right),
\end{eqnarray}
that does not affect the density $|\Psi({\bf r},t)|^2$. Indeed, we can make this point obvious in Eq.~(\ref{Eq:basic1}) by shifting the energy variable so that it will correspond to the kinetic energy instead of the total energy, $n({\bf r},\varepsilon,t)=\tilde{n}({\bf r},\varepsilon+\vartheta({\bf r},t),t)$. Expressed in the new coordinates the equation reads
\begin{eqnarray}
\partial_{t}n({\bf r},\varepsilon,t)-\left[\nabla-\nabla\vartheta_{{\bf r},t}\partial_{\varepsilon}\right]D_{\varepsilon}\left[\nabla-\nabla\vartheta_{{\bf r},t}\partial_{\varepsilon}\right]n({\bf r},\varepsilon,t)=\delta(t)\;F(\varepsilon,{\bf r})\quad\label{Eq:basic2}.
\end{eqnarray}

An equation for the density $n({\bf r},t)$ can be obtained by integrating Eq.~(\ref{Eq:basic2}) in $\varepsilon$
\begin{equation}
\partial_{t}n({\bf r},t)-\frac{\tau}{m}\nabla^2\left(\overline{\varepsilon}({\bf r},t)+\lambda n^2({\bf r},t)\right)=\delta(t)\;n({\bf r},0),\label{Eq:basic3}
\end{equation}
where
\begin{eqnarray}
\overline{\varepsilon}({\bf r},t)=\int \frac{d\varepsilon}{2\pi}\;\varepsilon n({\bf r},t,\varepsilon)\equiv \varepsilon_{kin}({\bf r},t) n({\bf r},t).
\end{eqnarray}
It can be written in the compact form $\partial_{t}n-\nabla^2(D_{ef\!f}n)=\delta(t)n$ when defining an effective space and time-dependent diffusion coefficient $D_{ef\!f}=(\varepsilon_{kin}+\lambda n)\tau/m$. The apparent simplicity of these equations is however deceiving. They are not closed equations for the density evolution, since the kinetic energy $\overline{\varepsilon}$ depends on the nonlinearity and needs to be determined separately via Eq.~(\ref{Eq:basic2}). Nevertheless, we arrive at the conceptually important result that in the diffusive regime the nonlinearity effectively introduces a density dependence of the diffusion coefficient. It seems clear that a closed form solution of the nonlinear equations for arbitrary initial conditions cannot be found. In order to make progress we will rely on two approaches: the use of conservation laws and the study of solvable limiting cases. When combined, they will enable us to arrive at a qualitative picture both for repulsive and attractive nonlinearity.

\begin{figure}[t]
\centerline{\includegraphics[width=4cm,angle=270,clip=]{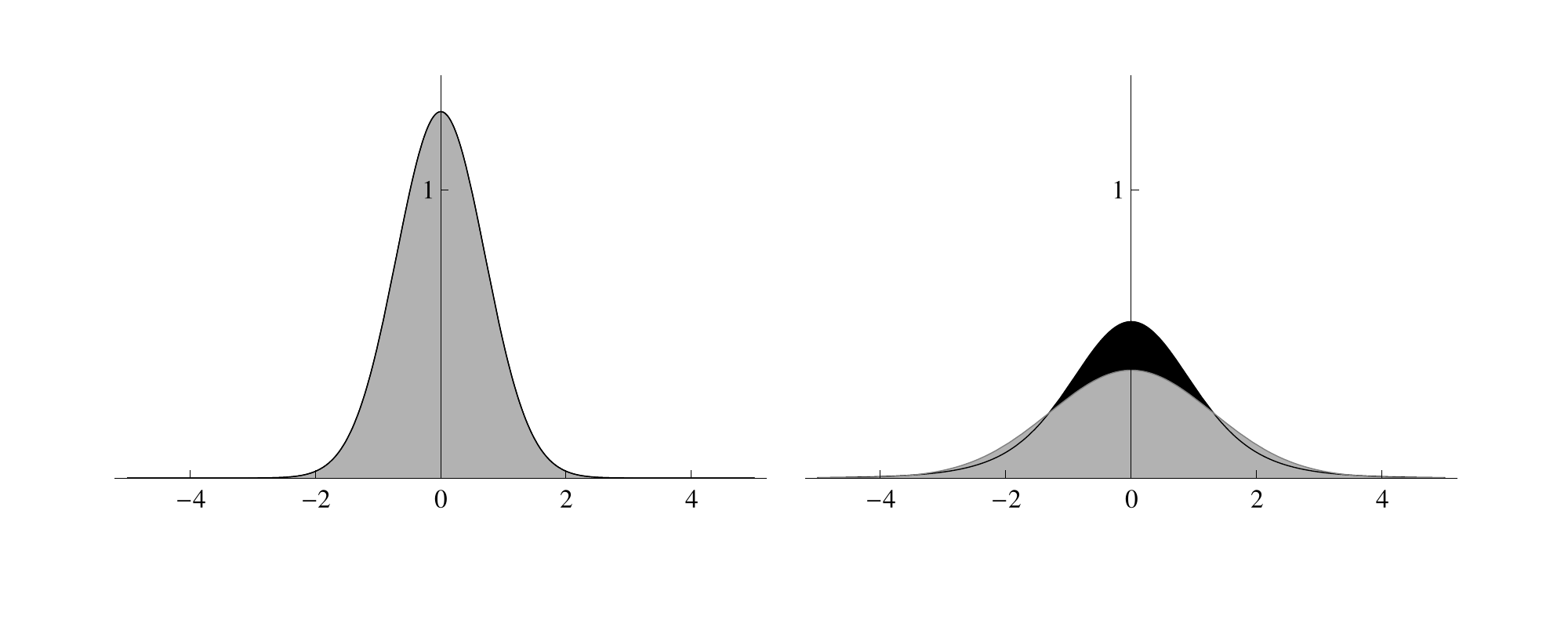}}
\caption{This figure compares two different \emph{linear} diffusion processes, one with a fixed diffusion coefficient (gray) and one with \emph{energy}-dependent diffusion coefficient $D_\varepsilon$ and an initial energy distribution of the form $C \exp(-\varepsilon/\varepsilon_0)$, where $C$ and $\varepsilon_0$ are constants. On the left hand side the initial density distribution is shown, which is chosen to be identical for both processes. After a while, the density that corresponds to energy dependent diffusion is larger both for small distances and for large distances from the origin when compared to the diffusion process with a fixed diffusion constant $D_{\varepsilon_0}$. This reflects the fact that both energies $\varepsilon\ll\varepsilon_0$ and $\varepsilon\gg \varepsilon_0$ are present in the initial energy distribution, leading to slow and fast diffusion respectively.}
\label{fig:diffvsdiff}
\end{figure}

First we briefly discuss the linear case, $\vartheta=0$. In the absence of nonlinearity, $n({\bf r},\varepsilon,t)$ evolves independently for each energy $\varepsilon$. In this limit, Eq.~(\ref{Eq:basic2}) has the obvious solution
\begin{eqnarray}
n({\bf r},{\varepsilon},t)=\frac{\Theta(t)}{4\pi D_{\varepsilon} t} \int d{\bf r}_1\;\mbox{e}^{-({\bf r}-{\bf r}_1)^2/(4D_{\varepsilon} t)} F({\varepsilon},{\bf r}_1).
\end{eqnarray}
Here, for each energy $\varepsilon$ diffusion is determined by the corresponding diffusion coefficient $D_{\varepsilon}$, and should be weighted according to the energy distribution in the injected wave-packet. The linear case was discussed in Ref.~\refcite{Shapiro07}. Starting from a broad energy distribution centered around some $\varepsilon_0$, both tails and central part of the density in the long time limit are more pronounced compared to diffusion at fixed energy ${\varepsilon_0}$, because the tails are determined by energies $\varepsilon>\varepsilon_0$, while the central part is dominated by energies $\varepsilon<\varepsilon_0$, see Fig.~\ref{fig:diffvsdiff}.

Next we turn to the nonlinear case. We will make use of the conservation laws for particle number and energy. By integrating Eq.~(\ref{Eq:basic3}) over ${\bf r}$, we obtain that the particle number (or normalization) $\int d{\bf r} \;n({\bf r},t)=N$ is fixed in time. An equation for $\overline{\varepsilon}({\bf r},t)$ can be derived by first multiplying Eq.~(\ref{Eq:basic2}) by $\varepsilon$ before integrating in this variable. Then by combining the equation for $n({\bf r},t)$ with the equation for $\overline{\varepsilon}({\bf r},t)$ we find that the energy
\begin{eqnarray}
E_{tot}=\int d{\bf r}\;\left(\overline{\varepsilon}({\bf r},t)+\lambda n^2({\bf r},t)\right)
\end{eqnarray}
is constant in time. Remarkably, this conservation law completely determines the time evolution of the mean radius squared of the wave-packet, $\left\langle r^2 \right\rangle \equiv \int d{\bf r}\;r^2\;n({\bf r},t)/N$. Indeed, multiplying Eq.~(\ref{Eq:basic3}) by ${r}^2$ and subsequently integrating in ${\bf r}$ one obtains that $\partial_t\left\langle{r}^2\right\rangle =4D_{\varepsilon_{tot}}$, where $\varepsilon_{tot}=E_{tot}/N$. The linear dependence of the mean square radius on time during the whole evolution is guarded by energy conservation. This is one of the central results of this paper. When compared to the linear case, the effective diffusion coefficient $D_{\varepsilon_{tot}}$ is reduced for attractive and enhanced for repulsive nonlinearities.

\begin{figure}[t]
\centerline{\includegraphics[width=4cm,clip=]{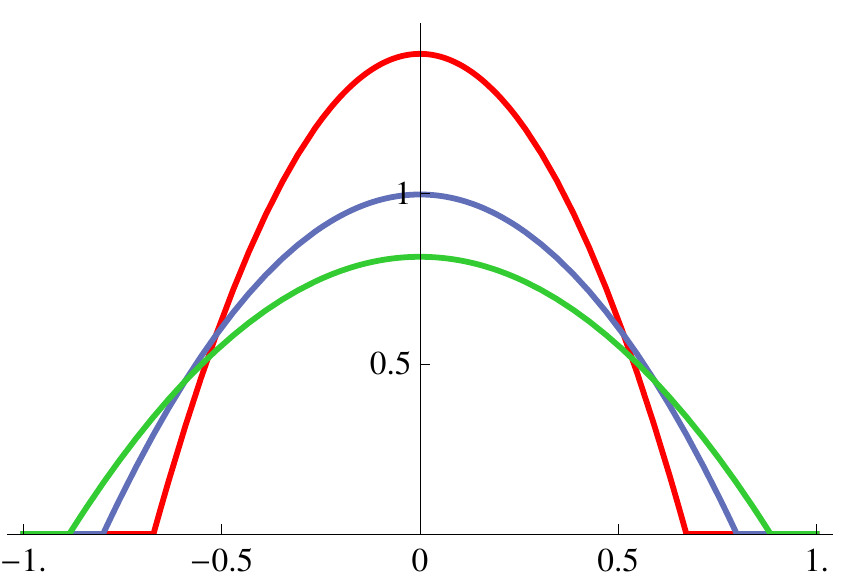}}
\caption{This figure shows the time-evolution of the Barenblatt-solution of the porous medium equation. It has the shape of an inverted parabola. It is worth noting that the wave-front has a finite spatial derivative at the boundary of the distribution. For this solution, the mean radius squared evolves rapidly as $\left\langle r^2 \right\rangle \propto t^{1/2}$ and correspondingly the density at $r=0$ drops as $n(0,t)\propto t^{-1/2}$.}
\label{fig:Barenblattbasicn}
\end{figure}

In the following we discuss more specifically the repulsive and attractive cases.
For the repulsive nonlinear case it is instructive to consider a situation in which the second term on the RHS of Eq.~(\ref{Eq:basic3}) dominates. The equation $\partial_tn=\nabla^2n^2$, which one obtains after simple rescaling, is an example of the famous porous medium equation (PME)\cite{Vasquez06}. For the $2d$ case the solution describing the evolution of a delta-function pulse $M\delta({\bf r})$ is given by $n({\bf r},t)=(C-r^2/(16t^{1/2}))/t^{1/2}$, where  $C^2=M/(8\pi)$\cite{Zeldovich50,Barenblatt52}. This solution is often referred to as Barenblatt's solution. It conserves the normalization $\int d{\bf r}\; n({\bf r},t)=M$ but, unlike ordinary diffusion, it is nonzero only in a finite region of space, see Fig.~\ref{fig:Barenblattbasicn}. The special importance of Barenblatt's solution in the theory of the PME is related to the fact that, roughly speaking, any solution starting from a sufficiently benign initial pulse with weight $M$ is eventually well-approximated by Barenblatt's solution with the same weight \cite{Vasquez06}.

\begin{figure}[t]
\centerline{\includegraphics[width=4cm,angle=270,clip=]{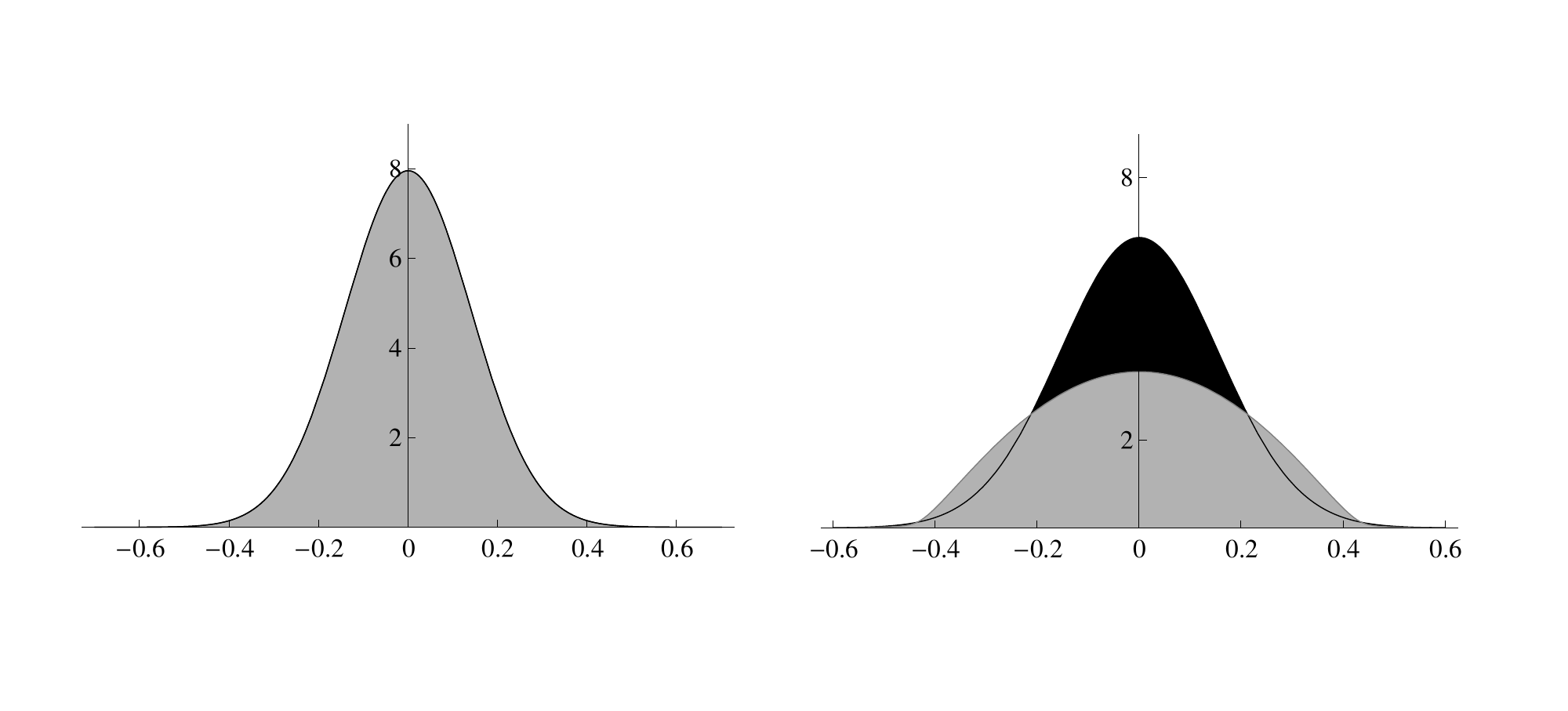}}
\caption{This figure compares a conventional diffusion process (black) to the time evolution described by the porous medium equation (gray) in two spatial dimensions. The same initial distribution is chosen for both processes and displayed on the left hand side. One can see, that the density at the center drops much faster for the solution of the porous medium equation. Simultaneously, the mean radius grows much faster, so that the total density is conserved.}
\label{fig:diffvspmeblack}
\end{figure}

\begin{figure}
\centerline{\includegraphics[width=7.5cm,angle=270,clip=]{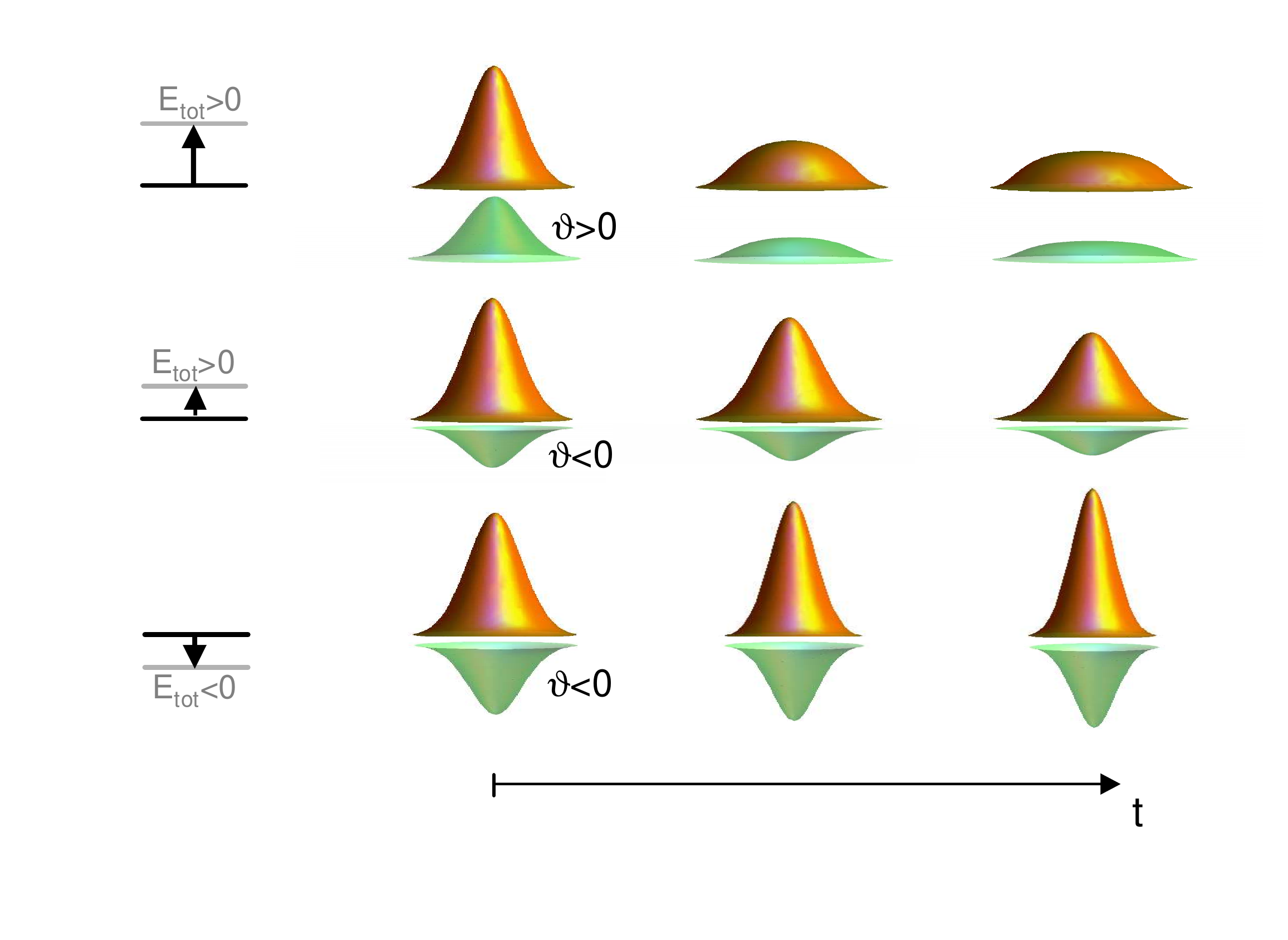}}
\caption{Time-evolution of the disorder averaged density $n$ sketched qualitatively for the three relevant cases. According to the relation $\partial_t\left\langle r^2\right\rangle =E_{tot}\tau/m$, the radius grows for $E_{tot}>0$ and shrinks for $E_{tot}<0$. The condition $E_{tot}>0$ can be realized for positive (first line) or negative potential $\vartheta=\lambda n$ (second line), while $E_{tot}<0$ can be realized only for negative $\vartheta$ (third line). In the repulsive case a rapid drop of the density distribution is expected in the center for large $\vartheta=\lambda n$, while $\left\langle r^2 \right\rangle$ grows only linearly in $t$ ("locked explosion"). For $E_{tot}<0$ a "diffusive" collapse is expected. In the cases with $\vartheta<0$ a fragmentation of the cloud may occur. The evolution is determined by the initial density \emph{and} energy distributions.}
\label{fig:2}
\end{figure}

For Barenblatt's solution, the mean radius squared evolves as $\left\langle r^2 \right\rangle \propto t^{1/2}$ and the density at $r=0$ drops as $n(0,t)\propto t^{-1/2}$. At short times this solution describes a much faster "explosive" evolution than the source-type solution of the diffusion equation, for which $\left\langle r^2\right\rangle \propto t$ and $n(0,t)\propto t^{-1}$, see Fig.~\ref{fig:diffvspmeblack}.
At first sight there seems to be a contradiction. If one injects a bell-shaped pulse with a large potential energy, it appears that the potential part of the effective diffusion coefficient $D_{ef\!f}=(\varepsilon_{kin}+\lambda n)\tau/m$ dominates. Therefore, naively, one would assume that the initial evolution is "explosive", while our exact result $\left\langle r^2 \right\rangle=r_0^2+4D_{\varepsilon_{tot}}t$ rules out this possibility. This puzzle can be resolved in the following way. The explosion takes place only in the central part of the density distribution, which has only a small weight when calculating $\left\langle r^2 \right\rangle$. Right in the center, for ${\bf r}=0$, Eq.~(\ref{Eq:basic3}) can be written as
\begin{eqnarray}
\partial_tn=\frac{\tau}{m}\left[\nabla^2\overline{\varepsilon}+2\lambda n\nabla^2n\right]
\end{eqnarray}
for $t>0$; we consider here a rotationally symmetric distribution with $\nabla n(0,t)=0$ and $\nabla^2n(0,0)<0$. For sufficiently large $\lambda n$, the potential part is dominant and leads to a fast initial decrease of the density before either $\lambda n\nabla^2 n$ becomes small or $\nabla^2\overline{\varepsilon}$ becomes positive as a consequence of the outward-flow of the kinetic energy. Away from the center, where the density and correspondingly the term $2\lambda n\nabla^2n$ are small, Eq.~(\ref{Eq:basic3}) takes the form
\begin{eqnarray}
\partial_tn\approx\frac{\tau}{m}\left[\nabla^2\overline{\varepsilon}+2\lambda (\nabla n)^2\right]
\end{eqnarray}
for $t>0$. For the PME it is the second term that determines the propagation of the boundary. For Eq.~(\ref{Eq:basic3}), however, the large kinetic energy outside the center leads to $\nabla^2\overline{\varepsilon}<0$ for intermediate distances, and this prevents the term $2\lambda(\nabla n)^2$ from dominating. It is therefore an inversion of the distribution of kinetic energy compared to that of the density that does not allow for an explosive expansion and leads to a linear dependence of $\left\langle r^2\right\rangle$ on $t$. A sketch of a typical density evolution expected for this "locked explosion" is presented in the first line of Fig.~\ref{fig:2}.

We now discuss general features of wave-packet dynamics in the disordered and nonlinear medium \cite{Hartung08}. For an expanding wave-packet, i.e. $E_{tot}>0$, the overall potential energy related to the nonlinearity is converted into kinetic energy. As a result, the total kinetic energy increases in the repulsive case and decreases in the attractive case. Correspondingly, during the course of the expansion localization effects can be expected to be weakened for repulsive nonlinearity and enhanced for attractive nonlinearity. In particular, for an attractive (self-focusing) nonlinearity the slowing down and eventual localization of the injected pulse (not considered here) occurs at smaller distances than in the linear case as observed in the experiment \cite{Schwartz07}. [Regarding the role of interference effects for localization three main factors need to be accounted for. As long as the evolution does not come to a halt, the time-dependent potential leads to dephasing, which weakens localization effects. When the wave-packet becomes broader, longer paths become available for interference and at the same time the kinetic energy decreases, whereby $\varepsilon\tau$ decreases. The latter two effects support localization.]

The attractive case is richer than the repulsive one (see Fig. 2), because the total energy may also be negative, $E_{tot}<0$. Then the mean radius squared would become equal to zero after a finite time. This corresponds to a celebrated phenomenon in nonlinear physics, the collapse \cite{Vlasov71,Sulem99}. Here it is realized for the diffusive system. To the best to our knowledge, this "diffusive" collapse has not been discussed in the literature. Since our reasoning is based on a diffusive kinetic equation and thus assumes frequent scattering, the linear decrease of $\left\langle r^2 \right\rangle$ only holds as long as the radius of the cloud exceeds the mean free path. [In the {\it clean case} the virial theorem for the NLSE in 2d \cite{Vlasov71,Sulem99} states that the {\it second} time derivative of $\left\langle r^2\right \rangle$ is proportional to the total energy. In this article we describe diffusive motion and correspondingly obtain a different time dependence for the size of the cloud.]

Even for $E_{tot}>0$ the collapse can play a role when the nonlinearity is attractive, if part of the cloud has a {\it negative energy}, while the remaining part expands. As a result one can expect a fragmentation of the cloud. If a part of the cloud with a {\it positive but small energy} lags behind, this fragment may have a strong tendency to localize. One may expect that this kind of localized or collapsing fragment generically remains from an expanding cloud with $E_{tot}>0$ but attractive nonlinearity.

To conclude, the nonlinear diffusion equation discussed in this paper contains rich physics that invites further theoretical and experimental investigations.

\section*{Acknowledgements}
The research was supported by the Minerva Foundation. We thank H.~U.~Baranger, A.~Belyanin, C.~Di~Castro, G.~Falkovich, Y.~Lahini, C.~A.~M\"uller, V.~L.~Pokrovsky, Y.~Silberberg, J.~Sinova and M.~D.~Spector for their interest in the work.

\bibliographystyle{ws-rv-van}

\end{document}